# Text-to-Image Generation for Vocabulary Learning Using the Keyword Method

NUWAN T. ATTYGALLE*, University of Primorska, Slovenia and Université catholique de Louvain, Belgium
MATJAŽ KLJUN*, University of Primorska, Slovenia and Stellenbosch University, South Africa
AARON QUIGLEY, CSIRO Data61, Australia
KLEN ČOPIČ PUCIHAR*, University of Primorska, Slovenia and Stellenbosch University, South Africa
JENS GRUBERT, Coburg University of Applied Sciences and Arts, Germany
VERENA BIENER, University of Stuttgart, Germany
LUIS A. LEIVA, University of Luxembourg, Luxembourg
JURI YONEYAMA, Nara Institute of Science and Technology, Japan and Inria Rennes - Bretagne Atlantique, France
ALICE TONIOLO, University of St Andrews, United Kingdom
ANGELA MIGUEL, University of St Andrews, United Kingdom
HIROKAZU KATO, Nara Institute of Science and Technology, Japan
MAHESHYA WEERASINGHE*, University of Primorska, Slovenia

**The Keyword Method**

Word to-be-learnt in Portugese: *lago*
Meaning in English: *lake*

Keyword (familiarly-sounding word): *log*
Association (description of a memorable link) between the keyword and the word representing the meaning of the foreign word: *a log floating in the middle of the lake*

**Generated images from the keyword method description with text-to-image generator**

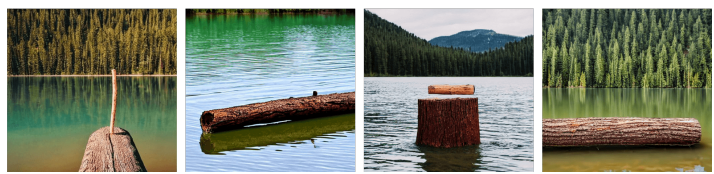

Fig. 1. The enhanced keyword method with visuals using text-to-image generation to reinforce word memorisation. Left: the explanation of the keyword method forming the link between the word to be learnt in the Portuguese language "lago" (meaning "lake" in English) and the similarly sounding word "log" (keyword). Right: four images of the description of the link (association) generated by DALL-E 2 text-to-image generator.

The 'keyword method' is an effective technique for learning vocabulary of a foreign language. It involves creating a memorable visual link between what a word means and what its pronunciation in a foreign language sounds like in the learner's native language. However, these memorable visual links remain implicit in the people's mind and are not easy to remember for a large set of words. To enhance the memorisation and recall of the vocabulary, we developed an application that combines the keyword method with

*Authors contributed equally to this research.

Authors' addresses: Nuwan T. Attygalle, University of Primorska, Koper, Slovenia and Université catholique de Louvain, Louvain-la-Neuve, Belgium, nuwan.attygalle@famnit.upr.si; Matjaž Kljun, University of Primorska, Koper, Slovenia and Stellenbosch University, Stellenbosch, South Africa, matjaz.kljun@upr.si; Aaron Quigley, CSIRO Data61, Eveleigh, Australia, aaron.quigley@csiro.au; Klen Čopič Pucihar, University of Primorska, Koper, Slovenia and Stellenbosch University, Stellenbosch, South Africa, klen.copic@famnit.upr.si; Jens Grubert, Coburg University of Applied Sciences and Arts, Coburg, Germany, jens.grubert@hs-coburg.de; Verena Biener, University of Stuttgart, Stuttgart, Germany, verena.biener@visus.uni-stuttgart.de; Luis A. Leiva, University of Luxembourg, Esch-sur-Alzette, Luxembourg, luis.leiva@uni.lu; Juri Yoneyama, Nara Institute of Science and Technology, Nara, Japan and Inria Rennes - Bretagne Atlantique, Rennes, France, juri.yoneyama@inria.fr; Alice Toniolo, University of St Andrews, St Andrews, United Kingdom, a.toniolo@st-andrews.ac.uk; Angela Miguel, University of St Andrews, St Andrews, United Kingdom, arm14@st-andrews.ac.uk; Hirokazu Kato, Nara Institute of Science and Technology, Nara, Japan, kato@is.naist.jp; Maheshya Weerasinghe, University of Primorska, Koper, Slovenia, maheshya.weerasinghe@famnit.upr.si.

text-to-image generators to externalise the memorable visual links into visuals. These visuals represent additional stimuli during the memorisation process. To explore the effectiveness of this approach we first run a pilot study to investigate how difficult it is to externalise the descriptions of mental visualisations of memorable links, by asking participants to write them down. We used these descriptions as prompts for text-to-image generator (DALL-E 2) to convert them into images and asked participants to select their favourites. Next, we compared different text-to-image generators (DALL-E 2, Midjourney, Stable and Latent Diffusion) to evaluate the perceived quality of the generated images by each. Despite heterogeneous results, participants mostly preferred images generated by DALL-E 2, which was used also for the final study. In this study, we investigated whether providing such images enhances the retention of vocabulary being learned, compared to the keyword method only. Our results indicate that people did not encounter difficulties describing their visualisations of memorable links and that providing corresponding images significantly improves memory retention.

**ACM Reference Format:**
Nuwan T. Attygalle, Matjaž Kljun, Aaron Quigley, Klen Čopič Pucihar, Jens Grubert, Verena Biener, Luis A. Leiva, Juri Yoneyama, Alice Toniolo, Angela Miguel, Hirokazu Kato, and Maheshya Weerasinghe. 2025. Text-to-Image Generation for Vocabulary Learning Using the Keyword Method. 1, 1 (January 2025), 24 pages. https://doi.org/10.1145/nnnnnnn.nnnnnnn

## 1 INTRODUCTION

Learning vocabulary is an important part of learning a foreign (target) language. A strong vocabulary enables us to read, write, listen, and speak with greater accuracy and fluency. The sooner we acquire a comprehensive set of words, the sooner we can engage in using the language [17, 72]. There are several learning methods and techniques available for learning vocabulary [47, 88]. Many of these methods and techniques rely on either providing an image as a visual stimulus (e.g. flash cards) or mentally visualising clues (e.g. learning words in context of a sentence, using a mnemonic such as a keyword method) related to foreign words. There is a large amount of evidence showing that creating mental visualisations as well as seeing an actual image (e.g. on paper or on screen), can significantly enhance recall. For example, seeing photos of past events results in significant recall of such events [25, 30, 35], reading a list of words and in addition imagining each word reduces false memories at recall time [15, 46, 66], and visiting a museum can result in greater recall of details about pictures even after a significant period of time [26].

With the aforementioned keyword method [2, 62] we need to visualise a memorable link (association) between what a word means and what its pronunciation in the foreign language reminds us of phonetically (keyword). For example, a Japanese word for a "book" to be learnt is "UTF8min", written in rōmaji as "hon". This sounds similar to "honey" (keyword) in English, while the visualised memorable link between the keyword and the word representing the meaning of the foreign word can be "honey dripping on a book" (association). Or the Portuguese word for "lake" to be learnt is "lago", which sounds familiar to "log" (keyword) in English and the visualised memorable link can be "a log in the middle of the lake" (association) (see Figure 1 left). When recalling a foreign word, one tries to remember the association, which reinforces bringing the word to mind. However, the method fully relies on mental visualisation only. We hypothesise that if these mental visualisations of associations could be externalised in actual images, additional visual stimuli could be provided to the user during memorisation and consequently reinforce recall. Thus, more words of a foreign language learnt could be remembered quicker, enabling faster fluency. However, images of the above mentioned associations were since lately very hard to come by.

Authors version.

The answer to this shortcoming can be synthetically generated images that merge together unrelated concepts with semantic coherence (such as "honey dripping on a book"). These machine learning text-to-image generators take a natural language description as an input and produce an image that matches that description. For the past couple of years text-to-image generators such as OpenAI's DALL-E[1], StabilityAI's Stable Diffusion[2] or Midjourney[3] are capable of producing high-quality realistic images that could be used for the purpose of externalising associations of the keyword method (see Figure 1 right). So far no previous work has explored this possibility.

We built a system to conduct three different studies investigating how text-to-image generators could enhance the keyword method for vocabulary learning in order to reinforce recall of the vocabulary being learnt. In the first pilot study, we explored whether users are capable of externalising their mental visualisations by using the keyword method in a text description, and if the images generated suitably represent these mental visualisations. In the second study, we compared how close different text-to-image generators come to what users consider a suitable representation of their mental visualisation. We used this study also to select the text-to-image generator for the final study. In the third and final study, we explored short and long term retention of foreign words in two conditions: (i) using mental visualisation of a memorable link only, and (ii) using mental visualisation and visual stimulus from the image generated by a text-to-image generator. The results show that people have no problems describing their visualisations of memorable links and that providing images of such links increases memory retention. Therefore, we conclude that providing images generated by text-to-image generator enhances vocabulary learning when used in combination with the keyword method.

## 2 RELATED WORK

Our research is structured around two primary areas: mnemonics-based learning techniques and text-to-image synthesis. In the following sections, we will delve into prior research in each of these domains.

### 2.1 Mnemonics and the keyword method

Research on memory and learning has shown that comprehension and recall can be supported by different types of instructional methods and techniques that can be used to process and store information [12]. Even in language learning, the evidence suggests that incidental vocabulary learning (e.g. while reading) is not particularly efficient if it is not accompanied with systematic vocabulary learning using strategies that reinforce words memorisation [34, 52]. Several learning strategies can be employed while learning vocabulary of a foreign language [17, 72]. While research suggests that cognitive strategies [71] (e.g. creating lists of words, word repetition, reading words in sentences, etc.) are most commonly employed [77], the strategies based on form (e.g. studying the spelling and pronunciation) and associative meanings of words (e.g. synonyms, related words, scales etc.) – i.e. mnemonic strategies [71] – significantly improve vocabulary breadth (how many words a person knows), and depth (how well a person knows these words) [88]. Mnemonics are just one type of instructional technique designed for enhancing memory and recall [41, 56, 58] and are good to be used in combination with other strategies to grasp the form, meaning and use of vocabulary [44]. Mnemonics connect what needs to be learnt to prior knowledge through the use of visual and/or acoustic cues and associations, thus encoding new information with something more accessible or meaningful to the user. Basic types of mnemonic techniques rely on the use of phonetic systems, rhyming words, acronyms, or key words [16, 27, 28, 38, 41, 58, 87].

---

[1] https://openai.com/research/dall-e
[2] https://stability.ai/
[3] https://www.midjourney.com/showcase

The so-called 'keyword method' requires learners to mentally visualise a memorable link or association between what the word to be learnt in a foreign language means and what its pronunciation in a foreign language sounds like [3]. Paivio and Desrochers [49, 50] argue that formation of mental images aids learning. According to Paivio's Dual Coding Theory (DCT), mental visualisation and verbal cues [50, 84] are functionally independent and processed differently and along distinct channels in the human mind when a person encodes information about a particular concept. Based on DCT, a concept "tree" can be thus stored as a word and as an image (derived from all sensory information), and it can be later retrieved from each channel individually or from both simultaneously. Experimental research has demonstrated that mental visualisation improves memory performance, strengthens memory retention, recall, and reduces false memories [64]. For example, mental visualisation of words on a list we are memorising (Deese–Roediger–McDermott paradigm or DRM) reduces false memories at the time of recall [46], or that forming mental visualisation while reading text increases the amount of content remembered over time [37].

It is the dual representation in mind that also makes the keyword method effective. The keyword method provides a powerful tool for words that have a high degree of "imageability" (e.g. *moon* is the word with high "imageability" since most people can easily mentally visualise it, while *truth* is the word with low "imageability" since it is not easy to mentally visualise it) [65], or for word pairs (what the word means and the familiarly-sounding word/phrase) between which the learner can form some kind of semantic link [13]. The important thing is that the visualised memorable link should clearly relate to the thing being memorised. For instance, the Japanese word for a postcard is "UTF8min", written in rōmaji as "hagaki", which in English sounds like "hug a key", and the visual link between the two can be imagined as "a picture of a hand hugging a key on a postcard".

An extensive body of literature has explored and proven the effectiveness of the keyword method in the memorisation of vocabulary [56, 57, 85] as well as when comparing it against other vocabulary learning strategies [2, 3, 62, 78, 80–82]. For example, a recent study compared generic model based instructions and keyword method based instructions for vocabulary learning in AR environments, and found that the keyword method outperforms the former in short-term, long-term retention as well as in learning efficiency [81]. In another study, researchers compared rehearsal, semantic mapping and the keyword method, and found that the keyword method outperformed the other two [68]. It has also been shown that the keyword method is superior to systematic teaching [33, 57]. Participants who were given a keyword along with the translation also remembered more words after 6 weeks [3].

Despite generally positive results, there have been some opposite findings when students have learning disorders [54] or when abstract terms are to be learned [76, 82]. The keyword method can thus be less effective when foreign words are not "keyword friendly" — that is, when they lack an obvious keyword or are difficult to visualise [20] or have low "imageability" [84]. This can be explained by concreteness effect stating that abstract ideas are difficult to imagine [7, 83]. "Concrete language is remembered better than abstract language on a variety of tasks" because concrete words can be processed as images as well as words [51, 83].

Nevertheless, the keyword method is still an effective technique to learn vocabulary. However, as many of the mnemonics, it solely relies on forming mental visualisation with no visual (or other) stimuli of the link between what the word means and the familiarly-sounding word/phrase. Using visuals besides text is recognised to motivate learners, assist the learning process, and increase the performance in a variety of subjects (e.g. mathematics [55], history [9], chemistry [10]) including language learning [39]. In vocabulary learning the actual pictures are often used to provide additional visual clues when learning new words [11] as depicted by a plethora of picture dictionaries (e.g. [1]) and methods, such as flash cards combining words and images. We hypothesise that forming mental visualisation of the link and externalising it into a visual stimulus using technology could thus provide additional benefits to the keyword method and thus further enhance memorisation and recall. In the past it was difficult to find pictures of mental visualisations such as "a picture of a hand hugging a key on a postcard". However, novel artificial intelligence text-to-image generators can be used to externalise such visualisation by generating photo-realistic synthetic images. In this paper we aim to investigate if combining the two can enhance the effectiveness of the keyword method.

## 2.2 Text-to-Image Synthesis

Automatically creating images that semantically match a user-provided text description is a challenging problem, which has proved elusive until the advent of deep learning, and more concretely until the advent of deep generative models. Before, attempts to build text-to-image generators were limited to simple collages of low quality (e.g. [89]). In an early work [40] researchers reversed the AlignDRAW algorithm [18] (originally designed for image captioning) to generate images from text. However, the quality of the resulting images was not good enough for real-world usage. In the last five years, different deep generative models have been devised to synthesise images from text, including Generative Adversarial Networks (GANs) [63], Variational Auto-Encoders (VAEs) [31], flow-based models [86], Autoregressive models [14], and more recently Diffusion models [67].

Diffusion models have recently demonstrated their ability to produce realistic images with simpler architectures. The key idea is to generate samples from the input data distribution first by adding random noise and then progressively removing the noise. Diffusion models can be slow, since many forward passes are required for denoising; however, less expensive alternatives have been developed to reduce the number of passes [75] and nowadays they are considered state-of-the-art approaches to text-to-image generation. Therefore, in the following we discuss diffusion models for image generation used in this paper in more detail.

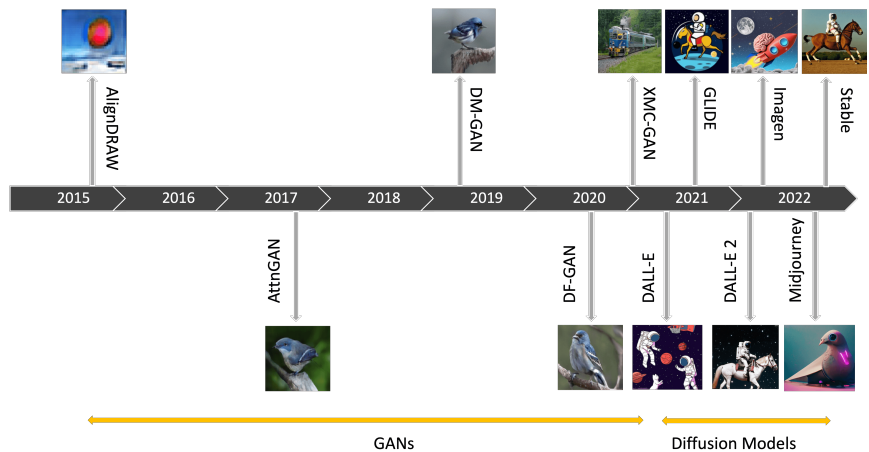

Fig. 2. Timeline of text-to-image synthesis milestones.

DALL-E [61] is a transformer architecture to autoregressively model the text and image tokens as a single stream of data. To reduce the context size of the transformer (up to a factor of 192) without a large degradation in visual quality, the researchers used a discrete VAE to compress 256 × 256 RGB images into a 32 × 32 grid of image tokens. The more recent version DALL-E 2 [60] uses a diffusion decoder to invert a CLIP image encoder [59]. Since the inverter is non-deterministic, DALL-E 2 can produce multiple images corresponding to a given CLIP embedding.

Midjourney[4] is another popular text-to-image generator based on diffusion models, developed by an independent company with the same name. Midjourney has gained traction particularly among visual artists, as it tends to generate painterly and surrealistic images [4]. One reason for its growing popularity is that Midjourney is accessible through a chatbot-like interface through Discord[5] servers.

Finally, Stable Difussion [67] is a new class of latent diffusion models that provides faithful and detailed reconstructions. The model is based on a CLIP-based image encoder, as in DALL-E, that provides a lower-dimensional representational space which was found to be perceptually equivalent to the data space. In contrast to previous work, the model is trained on the learned latent space, therefore it does not rely on excessive spatial compression. The reduced complexity also provides efficient image generation from the latent space with a single network pass.

Other diffusion models exist such as Imagen [69] and GLIDE [45], but their description goes beyond the aim of this paper. As described in this section, text-to-image generators work differently and can produce a variety of results. This is why one of our aims was to compare them in how close they come to users' mentally visualised links while using the keyword method.

## 3 THE OVERALL RESEARCH METHOD

The main research question of this work is whether it is possible to enhance the keyword method by externalising mental visualisation of associations with text-to-image generators to reinforce retention and recall. The question was further split into sub-questions: (RQ1) Are users capable to describe mental visualisation of associations formed by using the keyword method in a text description? (RQ2) Can text-to-image generators produce images that users would consider suitable? (RQ3) How good are different text-to-image generators in generating what users consider best representation of their associations? And the main research question (RQ4) How does the original keyword method compare to the one enhanced by actual externalised images?

In order to answer our research questions, we designed and developed a system for personal computers that combines the keyword method with a possibility to: (i) externalise the association between the meaning of the word and the familiarly-sounding word/phrase; (ii) externalise and appropriate the description of the mental visualisation of the link between the meaning of the word and the familiarly-sounding word/phrase; and (iii) generate visuals of the description by using text-to-image generator. Figure 3 shows the user interface of the system together with steps participants needed to follow for each word. Once the keyword and association were entered, the button to generate images became available and by clicking on it, the images were generated (with the association as a prompt) and shown. Among four (4) shown images, one could be selected. After all steps were completed, the participants could move to the next word.

Using our system we conducted three studies to answer research questions: Study A – A pilot study exploring the capability to externalise mental visualisation and suitability of generated image (to answer RQ1 and RQ2), Study B – Evaluating the ability of various text-to-image generators to externalise associations, identifying the one that users find most suitable in representing these associations, in order to use it in Study C (to answer RQ3). Study C – Comparing the classic keyword method to the enhanced version with externalised associations in the form of images (to answer RQ4).

For the pilot Study A the images were generated using DALL-E 2, because it was the latest one released and publicly available at the time. In Study B we compared the ability of different text-to-image generators to generate images users

---

[4]https://www.midjourney.com
[5]https://discord.com/

Authors version.

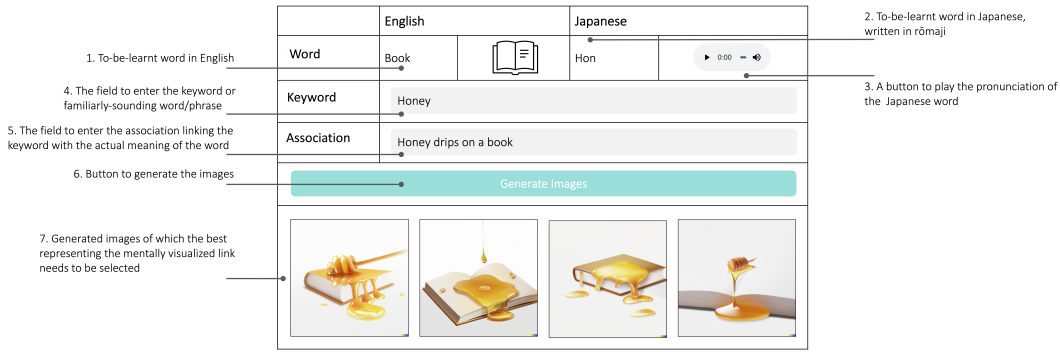

Fig. 3. System interface showing the (foreign) Japanese word for a book, a button for playing its pronunciation, fields for entering the keyword and association, and a button to generate images. Once the most suitable image is selected, users can move to the next word to be learnt in a foreign language.

find most suitable. Based on the results we used DALL-E 2 also for the main Study C. Due to licensing restrictions all image generations were done manually by the researcher who pushed the generated images into the interface. From the user perspective, the process looked automated. Each study is explained separately in the following sections. The data underlying this article will be shared on reasonable request to the corresponding author.

The university's committee for ethics in human subjects research has approved all three studies. Participants in studies have not been compensated.

## 4 STUDY A – EXPLORING THE CAPABILITY TO EXTERNALISE MENTAL VISUALISATION AND SUITABILITY OF GENERATED IMAGES

In this pilot study we explored if users are capable of externalising mental visualisations of associations they formed by using the keyword method and if state-of-the-art text-to-image generator can produce images that participants would consider suitable.

Participants were asked to remember five (5) pre-selected Japanese words (words with high "imageability" ratings from the MRC database [84] – see Appendix A) using the interface shown in Figure 3. The initial low number of words for this pilot study was selected in order to explore the difficulty of the task, when users are not familiar with the method. The participants were sitting at the desk with the keyboard, mouse and the screen in front of them. They were given a short explanation about the keyword method as well as a training session with two words before starting. For each of the five (5) words participants had to (i) read the word in English and Japanese, (ii) play the audio pronunciation of the word in Japanese, (iii) come up with and write down the keyword, (iv) come up with the association and write down its description in English, (v) click on the button to generate four (4) images, and (vi) select the one that best represents their association. Participants were then asked to rate on a 5-point Likert scale: (i) How difficult was it to come up with the association? (1 - very difficult, 5 - very easy), (ii) How satisfied were you with the generated image? (1 - very dissatisfied , 5 - very satisfied) and (iii) How much effort did you put into generating the association? (1 - a lot of effort, 5 - no effort).

After repeating this for five words, participants also completed the NASA Task Load Index (NASA TLX) [21, 43] to measure participants' subjective level of workload or *mental effort*. Next we measured the *immediate recall* by asking participants to remember all five Japanese words one by one by telling them the English word. If they could not remember the word, we showed them the selected generated image, and ask them again, to see if the image would trigger recall. We measured the immediate recall despite it not being part of the research questions in Study A (i) to test the method for Study C and (ii) to explore whether our interface is suitable for the learning task. Participants also filled in a short questionnaire with demographic questions. The entire experiment took around 20 minutes.

### 4.1 Participants

The study was completed by 10 participants, all voluntarily recruited, that consented to the study after reading an explanation about the procedure and data collection. They were also informed that they can abandon the study at any time. None of the participants had any prior knowledge of the Japanese language identified via a short competency test. In this test we showed 10 words for 10 common nouns, played their pronunciation, and asked participants to select the right answer for each word among three answers. The selection criteria was not knowing any of the words, which all participants met. They were all students at a European university on the English programmes (graduate level) in computer science related fields. This means that they were proficient in at least two (2) languages – their first language and English. The sample comprised of two (2) participants identified as female, while eight (9) identified as man. Participants were aged between 29 to 47, with a mean of $\bar{x}$ = 35.9 and standard deviation $\sigma$ = 7.2.

### 4.2 Data Collection and Analyses

Besides the questionnaires (for individual words, NASA TLX) and the *immediate recall* test mentioned, the *task completion time*, the time taken for the immediate recall test, and *image rendering time* were also logged by the system. The *image rendering time* was collected to determine waiting time.

The *learning efficiency* metric was first proposed in [48] and is calculated from task performance and task difficulty. In our case the performance was based on the *immediate recall* scores, and the difficulty on the *mental effort* participants invested in the learning phase determined by NASA TLX. First, the z-scores were calculated from performance and task difficulty scores separately (as $z_P$ and $z_M$ respectively) using the formula $z = (r - M)/\sigma$ where $r$ = Raw data score, $M$ = Population mean, and $\sigma$ = Standard deviation. Next, the *learning efficiency* was calculated using $E = (z_P - z_M)/\sqrt{2}$ For more details see [8, 19, 48].

To describe and summarise the characteristics and distribution of the values, we performed a descriptive statistical analysis by calculating the means ($\bar{x}$) and standard deviations ($\sigma$) for each data set collected in the study. To measure the reliability of the recall test, we conducted a Kuder-Richardson formula 20 test [36]. We found that $KR = 0.71 > 0.5$, indicating that the reliability of the recall test is acceptable.

### 4.3 Results of Study A

Descriptive statistics for the questions about the difficulty of association creation, image satisfaction, and effort invested in association creation are described in Table 1.

The results show that participants found it neither difficult nor easy (1 - very difficult, 5 - very easy) to come up with associations ($\bar{x}$ = 2.84, $\sigma$ = 1.33). The effort invested into generating associations (1 - a lot of effort, 5 - no effort) was also rated neither easy nor hard ($\bar{x}$ = 2.86, $\sigma$ = 1.23). The result for *mental effort* (NASA TLX, $\bar{x}$ = 39.40, $\sigma$ = 20.96) also fell in the middle of the range (0-9 is "low", 10-29 "medium", 30-49 "somewhat high", 50-79 "high", and 80-100 "very high"). An interesting observation was that users used keywords (familiarly sounding words) mainly in their first language (not English), but they also used similarly sounding words from other languages they spoke, (geographical

|  | Mean | SD | 95% CI Lower | 95% CI Upper | Min | Max |
|---|---|---|---|---|---|---|
| **Difficulty of Association Creation** (1 - very difficult, 5 - very easy) | 2,84 | 1,33 | 2,46 | 3,22 | 1,00 | 5,00 |
| **Image Satisfaction** (1 - very dissatisfied , 5 - very satisfied) | 4,10 | 0,97 | 3,82 | 4,38 | 1,00 | 5,00 |
| **Effort for Association Creation** (1 - a lot of effort, 5 - no effort) | 2,86 | 1,23 | 2,51 | 3,21 | 1,00 | 5,00 |
| **Immediate Recall** (percentage) | 80,00 | 31,30 | 57,60 | 102,40 | 0,00 | 100,00 |
| **Mental Effort** (0-9 is low, 10-29, 30-49, 50-79, 80-100 very high) | 39,40 | 20,96 | 24,41 | 54,39 | 8,00 | 72,00 |
| **Immediate Learning Efficiency** (bigger positive - HE*, lower negative - LE*) | 0,79 | 0,92 | 0,13 | 1,45 | -0,78 | 2,50 |
| **Task Completion Time** (minutes) | 3,38 | 1,55 | 2,94 | 3,82 | 1,46 | 8,96 |
| **Image Rendering Time** (minutes) | 0,80 | 0,31 | 0,72 | 0,89 | 0,23 | 1,78 |

*HE - high efficiency, LE - low efficiency

Table 1. Descriptive statistics for Study A. Note that the confidence interval (CI) of the mean assumes that sample means follow a t-distribution with $N − 1$ degrees of freedom.

and personal) names, etc. Nevertheless, the association descriptions had to be written in English for text-to-image generation to work, which was not a problem.

The satisfaction level with how images suitably represented participants' mental visualisation of their description was high ($\overline{x}$ = 4.10, $\sigma$ = 0.97). *Task completion time* for each word (from step (i) to step (vi)) was $\overline{x}$ = 3.38 or 3:23 minutes ($\sigma$ = 1.55 or 1:33 minutes). The *image rendering time* was also short enough for not making participants wait.

In *immediate recall* test, 80% of all words were remembered immediately and additional 6% were remembered (or 30% of 20% not remembered) after the image generated was shown. One participant was an outlier – they did not recall any of the words before and only one after images were shown. Excluding this participant, 88.9% or words were remembered immediately and 4% (or 40% out of 11.1% not remembered) after the image generated was shown. The *learning efficiency* score was $\overline{x}$ = 0.79 ($\sigma$ = 0.92), which places it in the "high efficiency" area.

The study showed promising results to continue with the second study.

## 5 STUDY B – EXPLORING THE CAPABILITY OF DIFFERENT TEXT-TO-IMAGE GENERATORS TO SUITABLY EXTERNALISE MENTAL VISUALISATIONS

Study B was designed to (i) compare how close different text-to-image generators come to what users consider a suitable representation of their mental visualisations, and (ii) select the best text-to-image generator for Study C. In the first phase participants had to come up with keywords and associations for 10 Japanese words (words with high "imageability" ratings from the MRC database [84] – see Appendix A) one after another. After they completed this for all 10 words they took a break. During this break one researcher talked to participants, while the other generated 16 images for each association using four different text-to-image generators (i.e. generating 4 × 4 = 16 images per association) using DALL-E 2[6], Stable Diffusion[7], Midjourney[8], and a generic latent diffusion model[9]. We selected these models because they were the newest publicly available for testing at the time of the study and were also compared in other studies (e.g. [4]).

---
[6]https://openai.com/index/dall-e-2/
[7]https://github.com/replicate/cog-stable-diffusion
[8]https://docs.midjourney.com/docs/early-models
[9]https://github.com/CompVis/latent-diffusion

The image generation process took approximately 5-10 minutes to complete after which participants were asked to select the best image for each word. The selection was made in 5 iterations. For each of the 10 words, in the 1st, 2nd, 3rd, and 4th iteration four (4) images were shown – one from each model. In each iteration participants had to select the most suitable image (i.e. the one that best fit their mental visualisation of the association). In the 5th and final iteration, four (4) previously selected images from each previous iteration were shown and participants had to again select the most suitable one (i.e. the overall winning image). After the final selection, users' were asked to fill in the questionnaire measuring appeal (e.g. attractiveness, desirability), hedonic (e.g. novelty, originality), and ergonomic (e.g. simplicity, familiarity) properties [22, 23] (used in software and product evaluation) with an open-ended question about why the overall winning image was selected.

### 5.1 Participants

We recruited 14 participants aged between 20 and 47, five (5) identified as female and nine (9) as man, eight (8) were students, and six (6) employed. All but one participant has a computer science background. Since this study did not measure the retention and recall, participants from Study A were also asked to participate. Participants were aged between 20 to 47, with a mean of $\bar{x} = 36.3$ and standard deviation $\sigma = 10.7$.

### 5.2 Data Collection and Analyses

Each participant performed the best image selection 50 times (10 associations $x$ (4 iterations + final winner selection)). In theory 700 image selections should have been recorded. However, for several associations we failed to generate all images due to content generation restrictions imposed by text-to-image providers. This happened despite instructions to avoid offensive or sensitive content. In total 640 image selections were recorded for 128 word and association pairs.

### 5.3 Results of Study B

The results in Figure 4 show that DALL-E 2 was selected in more than 40% in the first four (4) iterations, followed by Midjourney, Stable Diffusion and Latent Diffusion respectively. Looking at the final selection ratio (the 5th iteration), we can observe an even higher advantage of DALL-E 2 over other text-to-image generators (approximately 60% selection ratio) and again followed by Midjourney, Stable Diffusion and Latent Diffusion respectively.

Despite the clear advantage of DALL-E 2, users still selected images from other generators during the first four iterations of image selection ($\bar{x} = 1.9$, $SD = .76$). Furthermore, two thirds (66.67%) of images in the 5th iteration (final selection image sets) included images generated with two or more image generators. We hypothesise that sets containing images generated with different models result in higher diversity of imagery. To demonstrate this we sorted the collection of all final image sets based on the number of different text-to-image generators used. The two image sets in Figure 5 highlight this difference.

The results in Figure 6 provide some reasoning why users preferred certain image generators. As can be seen DALL-E 2 outperformed other generators for both ergonomic and hedonic qualities. However, it came second best in image appeal. The results also show large standard deviation across all metrics and generators (see Appendix B). Based on the results, we decided to use DALL-E 2 as the text-to-image generator in the main Study C.

Despite increasing the number of words to 10, participants who were not familiar with the keyword method did not find it difficult to come up with keywords and associations. For this reason we decided that we can increase the number of words for Study C as in similar studies where keywords and associations were pre-generated by researchers and provided to users [79, 80].

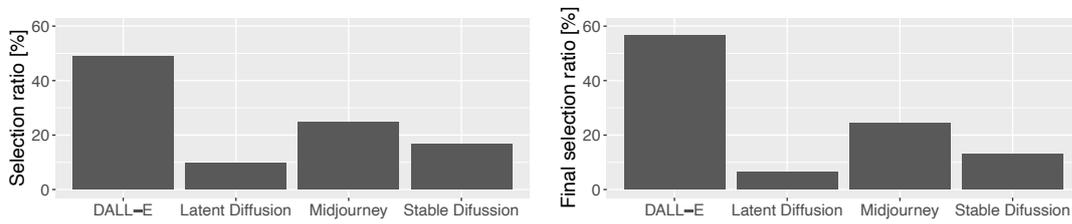

Fig. 4. Left: Selection ratio for the first four (4) iterations of image selection. Right: Final selection ratio where users select the best image of all.

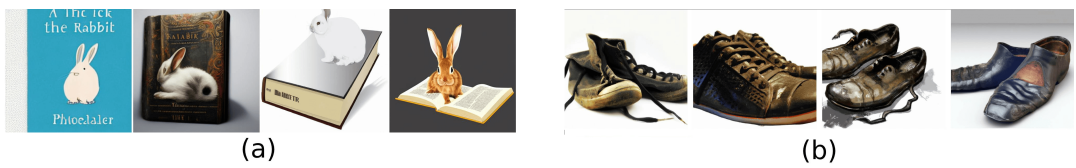

Fig. 5. (a) Final image set (for the 5th selection) generated using DALL-E 2 for text input "Usagi is holding a rabbit". (b) Final image set generated using four different image models for text input "bad shoes". From left to right: Latent Diffusion, Midjourney, DALL-E 2, and Stable Diffusion. Note: Images generated with different models offer a higher degree of diversity.

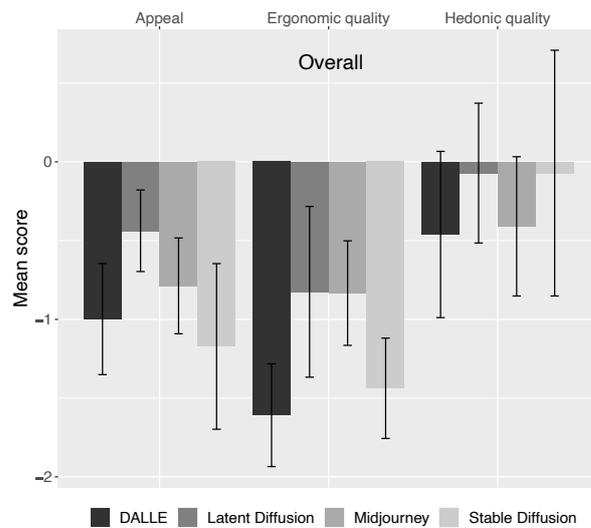

Fig. 6. Overall scores assessing appeal (e.g. attractiveness, desirability), hedonic qualities (e.g. novelty, originality) and ergonomic qualities (e.g. simplicity, familiarity) of generated images for each text-to-image generator.

## 6 STUDY C – COMPARING THE KEYWORD METHOD TO THE ENHANCED KEYWORD METHOD

In this study we compared two conditions: the keyword method that we call association, and the enhanced method we call association+visual condition. In each condition participants were asked to remember ten (10) pre-selected words (words with high "imageability" ratings from the MRC database [84] – see Appendix A) in a foreign language – one group learning Japanese and one Slovenian (between subjects factor) in order to investigate if there is an effect between different languages. This means that each person learnt 10 words (Set 1) in one and another 10 words (Set 2) in the other condition to account for individual differences. The sets of words were counterbalanced between conditions and the order of words in each set was randomised for each participant. The procedure was the same as in Study A, except that in the ASSOCIATION condition participants skipped generating (step (v)) and selecting the image (step (vi)).

After finishing with all ten words in one condition, completing the NASA Task Load Index (NASA TLX), and the *immediate recall* test (as in Study A participants were asked to remember Japanese words without and with the selected image if the former failed), participants were given a 10 minute break after which they repeated the procedure with the other condition (within subjects factor) with another set of 10 foreign words. To avoid the "order effects" [73], the order of the conditions as well as the order of the word sets were counter balanced.

After finishing the experiment, participants were given a post-questionnaire to evaluate their preference of the two conditions. The post-questionnaire had the following questions: (i) Which method did you find more efficient (you obtained better results)? (*retrieval efficiency*), (ii) Which method did you find easier to use? (*method difficulty*), (iii) Which method did you prefer? (*method preference*), (iv) How would you rate the method with images? (on a 3-point Likert scale from easy to hard), and (v) How would you rate the method without images? (on a 3-point Likert scale from easy to hard).

At the end, participants filled in a short questionnaire with demographic questions. The entire experiment took 90 to 100 minutes. After seven (7) days, participants completed the *delayed recall* test for all 20 words again in the same way as with *immediate recall*.

### 6.1 Participants

The study was completed by 32 participants, all voluntarily recruited. Of these 16 participated in learning Japanese and had no prior knowledge of the Japanese language identified via a short competency test. In this test we showed 10 words for 10 common nouns, played their pronunciation, and asked participants to select the right answer for each word among three answers. The selection criteria was not knowing any of the words, which all participants met. The participants were all students at a European university on the English programmes (undergraduate and graduate levels) in computer science related fields. This means that they were proficient in at least two (2) languages – their first language and English. The Japanese language was selected as a target language because it is from a distinct language family (all participants spoke Indo-European languages) and geographically far away.

The other 16 participants participated in learning Slovenian language and had no prior knowledge of any Slavic language also identified via a short competency test. These participants were graduate students in computer science at a Japanese university studying and working in the English speaking laboratory. They all passed English as a second language proficiency test with high scores in order to enrol to the university. The Slovenian language was selected because it is spoken by only 2 million people, which reduced the possibility of participants being familiar with it. As a south Slavic language it is also very different from English, which belongs to Germanic languages within the Indo-European family.

All participants consented to the study after reading the explanation about the procedure and data collection. They were also informed that they can abandon the study at any time. The sample comprised of nine (9) participants who identified as female (8 in the Japanese group and 1 in Slovenian learning group). The rest identified as male. All participants were aged between 19 to 47 years, with a mean of $\bar{x} = 25.2$ and $\sigma = 4.9$.

## 6.2 Data Collection and Analyses

A questionnaire for each word, data for *mental effort* (NASA TLX), *immediate recall* (percentage of words remembered), *delayed recall* (percentage of words remembered), *task completion time*, and *learning efficiency* were collected or calculated. *Learning efficiency* was calculated for *immediate recall* and *delayed recall* as in Study A. In addition, participants answered the post-questionnaire with five (5) questions as explained above.

Each dataset collected was first checked for normality using the Shapiro–Wilk normality test [74]. In all statistical analyses we used a significance level $p-value > 0.05$ and a restrictive confidence interval (CI) of 95%. For *immediate recall*, *delayed recall*, *mental effort*, *task completion time*, *immediate learning efficiency* and *delayed learning efficiency*, the statistical significance was examined using a Robust ANOVA test[10]. Asterisk notation is used in tables to visualise statistical significance (no significance or ns: $p > .05$, *: $p < .05$, **: $p < .01$, and ***: $p < .001$).

To measure the reliability of recall and preference questionnaires, we conducted a Kuder-Richardson 20 test [36]. Since the value is $KR = 0.65 > 0.5$, we can conclude that the reliability of the questionnaires is acceptable.

## 6.3 Results of Study C

*6.3.1 Method Preference, Retrieval Efficiency and Difficulty.* The results from the *method preference*, *retrieval efficiency* and *method difficulty* data from the post-questionnaire are shown in Figure 7. The majority of participants (66%) preferred the ASSOCIATION+VISUAL condition, and indicated that this condition helped them retrieve the words faster (50% of participants) and easier (47% of participants). As in Study A, we observed associations in several languages – mainly the first language (66%). Only 28% association used English whereas 6% mixed English with other languages (see examples in Appendix C).

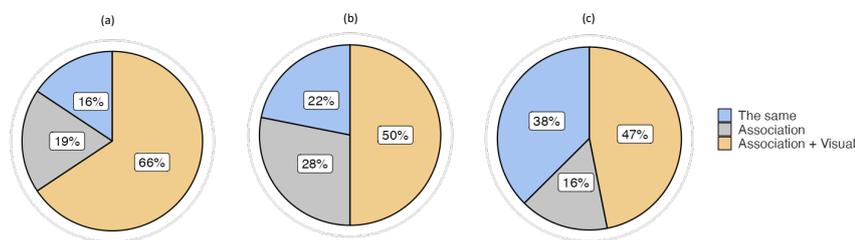

Fig. 7. The participants' answers to the questionnaire about method preference: (a) *method preference* – Which method did you prefer? (The same, Association, Association + Visual) ; (b) *retrieval efficiency* – Which method did you find more efficient?; and (c) *method difficulty* – Which method did you find easier to use?.

*6.3.2 Immediate and Delayed Recall.* The percentages of words remembered during *immediate* and *delayed recall* are presented in Table 2. The data shows that during *immediate recall* participants remembered more words in ASSOCIATION+VISUAL condition (86.93%) compared to the ASSOCIATION condition (81.86%). When providing help (a trigger) in the form of the description of the memorable link in the ASSOCIATION condition, and the image in the ASSOCIATION+VISUAL condition, participants remembered around 6% additional words in both conditions. The results for *immediate recall* are very similar to the results in Study A.

---

[10]Robust ANOVA, Dennis Twesmann, RPubs by RStudio, 07 Mai 2022 https://rpubs.com/DeTwes/robANOVA

In the *delayed recall* test participants also remembered significantly more words in the ASSOCIATION+VISUAL condition. however, the percentage of recalled words was lower (42.50% for ASSOCIATION+VISUAL and 29.06% for ASSOCIATION) compared to *immediate recall*. When providing help (a trigger) participants remembered around 15% additional words in the ASSOCIATION+VISUAL condition and around 22% for ASSOCIATION). The results are still in favour of ASSOCIATION+VISUAL condition with 57.32% compared to 51.58% of remembered words for ASSOCIATION condition.

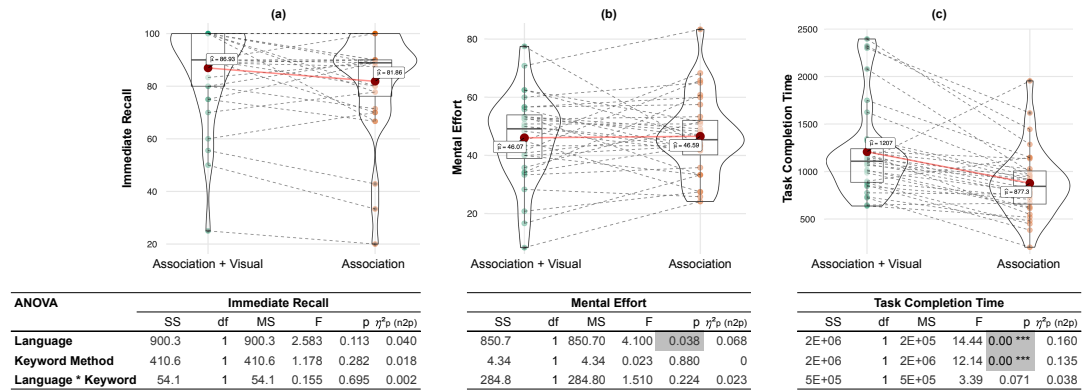

Fig. 8. The violin plots show the distribution of values for all users and probability density while the tables contain ANOVA results for: (a) *immediate recall* performance in percentage of correctly remembered words; (b) *mental effort*; (c) *task-completion-time* in seconds. In the violin plots the dashed grey lines link each user's results in both ASSOCIATION+VISUAL and ASSOCIATION condition. If the line is darker, more than one user had the same results. The red line links the mean values $\hat{\mu}$ between conditions. In tables the columns are $SS$ = sum-of-squares, $df$ = degrees of freedom, $MS$ = mean squares, $F$ = F-ratio, $p$ = p value, and $\eta_p^2$ = partial eta-squared for effect size. The first column in each table describes whether the ANOVA presents the results for the the effect of language (Japanese, Slovenian), keyword method condition (ASSOCIATION+VISUAL and ASSOCIATION), or the interaction effect of the two.

|  | Keyword Method | # of Participants | Recall | Recall with Help |
|---|---|---|---|---|
| Immediate Recall | Association + Visual | 32 | 86.93% | 92.21% |
|  | Association | 32 | 81.86% | 88.84% |
| Delayed Recall | Association + Visual | 32 | 42.50% | 57.32% |
|  | Association | 32 | 29.06% | 51.58% |

Table 2. The percentage of recalled words for all conditions.

The mean values of *immediate recall* and the ANOVA results across study conditions (i.e. the REPRESENTATION METHOD (ASSOCIATION and ASSOCIATION+VISUAL) and the LANGUAGE LEARNT (JAPANESE and SLOVENIAN)) are shown in Figure 8 (a). The mean values of *delayed recall* and the ANOVA results for study conditions are shown in Figure 9 (a). A significant main effect of the REPRESENTATION METHOD on *delayed recall* was detected ($F(1, 32) = .282$, $p < .05$, $\eta_p^2 = .077$). The language had no statistical effect on recall.

*6.3.3 Mental Effort and Task Completion Time.* The mean values of mental effort (measured by NASA-TLX) invested to carry out the learning task, are presented in Figure 8 (b). The ANOVA result indicated that there was no significant effect between the ASSOCIATION+VISUAL and ASSOCIATION ($F(1, 32) = .023$, $p > .05$, $\eta_p^2 < .01$). This was expected since additional generating of images was an easy sub-task composed of clicking on a button and selecting an image. However, language had a statistically significant effect size ($F(1, 32) = 4.1$, $p = .038$, $\eta_p^2 = .068$) with eta squared indicating a medium effect. The Japanese group learning Slovenian reported a higher mental effort ($\bar{x} = 50.0$, $\sigma = 12.3$) compared to the European group learning Japanese ($\bar{x} = 42.7$, $\sigma = 14.9$), which can be attributed to cultural differences as noted in other studies [32, 32]. However, since the language did not affect the recall, further investigation is beyond the scope of this paper and falls under future work.

The data for *task completion time* are shown in Figure 8 (c). A significant main effect between the ASSOCIATION+VISUAL and ASSOCIATION could be detected ($F(1, 32) = 12.14$, $p < .001$, $\eta_p^2 = .135$). Here, the *completion time* was significantly lower for ASSOCIATION condition ($\bar{x} = 877.31$ seconds or 14:37 minutes, $\sigma = 360$ seconds or 6 minutes) compared to the ASSOCIATION+VISUAL condition ($\bar{x} = 1207.13$ s or 20:12 minutes, $\sigma = 480$ s or 8 minutes). This was also expected since generating images took some time.

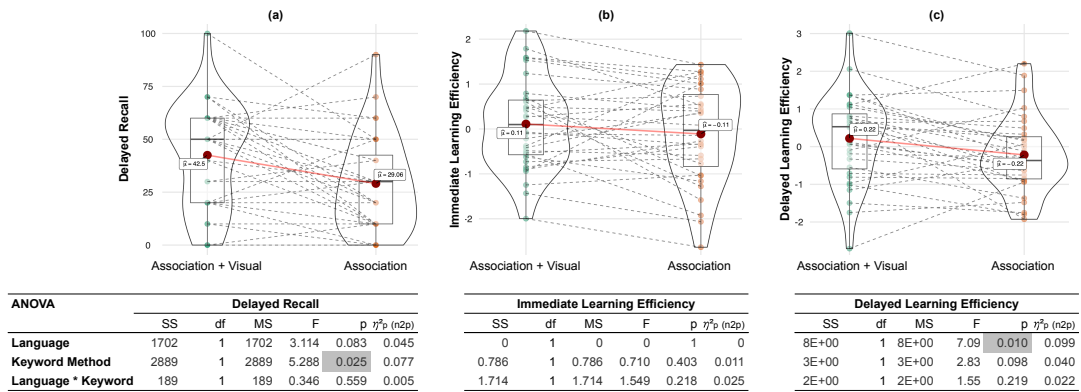

Fig. 9. The violin plots show the distribution of values for all users and probability density while the tables contain ANOVA results for: (a) *delayed recall* performance in percentage of correctly remembered words; (b) *immediate learning efficiency*; (c) *delayed learning efficiency*. The description is the same as in Figure 8.

*6.3.4 Learning Efficiency.* The mean *learning efficiency* values for *immediate recall* and *delayed recall* across study conditions are shown in Figure 9 (b) and Figure 9 (c). No significant effect could be detected between ASSOCIATION+VISUAL and ASSOCIATION on *learning efficiency* for *immediate recall* ($F(1, 32) = .403$, $p > .05$, $\eta_p^2 = .011$) and for *delayed recall* ($F(1, 32) = 2.83$, $p > .05$, $\eta_p^2 = .040$). However, a significant effect of the language on *delayed learning efficiency* was detected ($F(1, 32) = 7.09$, $p < .05$, $\eta_p^2 = .099$). The efficiency was significantly higher for the group learning Japanese ($\bar{x} = 0.346$, $\sigma = 1.074$) compared to the group learning Slovenian ($\bar{x} == -0.346$, $\sigma = 1.046$). Since efficiency depends on mental effort and performance and since language did not significantly affect the performance of the *delayed recall*, the significance is the result of the group learning Slovenian reporting a higher mental effort as discussed in subsubsection 6.3.3.

## 7 DISCUSSION

### 7.1 Implications, Design Recommendations, and future directions

The keyword method relies on mental visualisation of a memorable link between the meaning of the word and what its pronunciation in the foreign language reminds us of phonetically. In Study A we showed that users are able to describe their mental visualisation of memorable links into a text description (RQ1). We also showed that images produced by text-to-image generators were considered suitable (RQ2). Next, we compared the traditional method with the one where we synthetically generated images of the descriptions of memorable links. Although statistically significant for the delayed recall only, the results for both *immediate recall* and *delayed recall* show that participants recall more words when image was also generated and viewed during the study (RQ4). This was expected and is consistent with prior research showing that using visuals besides text motivates learners, assists the learning process, and increases performance in a variety of subjects [9, 10, 39, 55].

The results can be explained by the impoverished relational-encoding hypothesis that suggest that experiencing an image, focusing on its details and features, during the encoding process, reduces relational processing that typically activates false memories [24]. If images are observed by users, additional details of the image can be encoded and used during the recall time. As mentioned, the difference between the original and enhanced method grows in time. However, for both conditions the amount of words recalled during the *delayed recall* dropped significantly. While this was expected, it also suggests that if the method is to be used in computer software that takes the advantage of text-to-image generators, it should be combined with other methods that support continuous learning.

One example is combining it with the spaced repetition method in a spaced repetition software (SRS) [5, 6]. SRS are computer programs for memorising a list of items that can include vocabulary words. The foreign words appear on screen one by one in a sequential pattern and users try to memorise the words and indicate the difficulty of the foreign word by scoring it on a challenging scale. The software uses algorithms to space out intervals when each word will appear again in the future during the recall time. Harder to remember words indicated by users will appear sooner than easier to remember words, and harder to remember words that are mastered will thenn appear less often. Simple SR method could be augmented with the keyword method with images. When memorising a word, users could also describe the memorable link between the keyword and the actual meaning of the word as well as generate and select a suitable visual representation of the link. During the recall time, the system could show the visual image and description if needed to support the recall process.

An interesting result presented in the paper is the fact that triggering users by showing the image of the association compared to showing the text description of the association to help the recall seems to have no advantage. Users recalled the same percentage of words during *immediate recall* and even higher percentage for the condition without images during *delayed recall*. This should be explored further in the future.

Our Study B has shown that images generated by DALL-E 2 were selected most often. However, approximately 66% of images that made it to the 5th iteration (i.e. final selection set) were from at least two different models, showing that different models can also produce suitable images (RQ4). Generating images using different models increases the diversity of image sets. This diverse selection can be explained by the Von Restorff or "isolation" effect stating that when multiple homogeneous stimuli are presented, the one most likely to be remembered is the one that differs from the others [53]. It is likely that users selected the most distinctive image in each iteration resulting in a set of diverse images. Distinctive properties are also desirable while learning – more distinctive properties envisioned and integrated into memory during the encoding process, more likely is to recall these cues during the retrieval process [70].

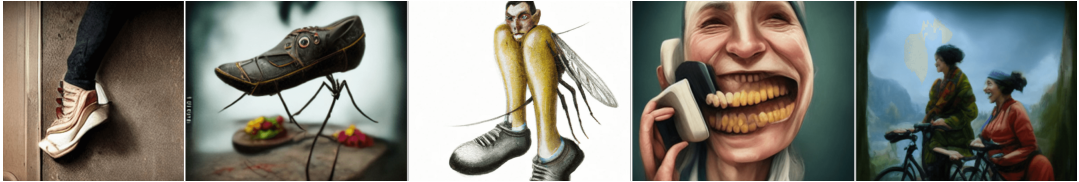

Fig. 10. Examples of absurd images participants selected as the final image in Study B.

In addition, some users descriptions of associations included bizarre or absurd combinations that can be seen in Figure 10. The idea that absurd images are easier to remember has been explored and confirmed in several studies (e.g. [42, 90]). However, others claim that users are more likely to recall previously seen multi-object scenes that are organised with novel but possible inter-object relations, as compared to organised scenes with familiar inter-object relations or even unorganised scenes with impossible inter-object relations [29]. As can be seen in Figure 10, the scenes in some of the generated images are unorganised and have impossible inter-object relations, yet they were selected in the final selection by users. The "isolation" effect and the effect of distinctiveness heuristic should be further explored in the future by comparing the recall of images generated by regular text-to-image prompts with the ones that take into account "distinctivness" and "novel possible inter-object relations".

One disadvantage of the keyword method is that it is hard to apply to abstract concepts, i.e. words that are difficult to visualise [20] or have low "imageability" [84]. In our studies we used terms for everyday objects and animals, and the words and their translations were composed of one word only. Potentially, abstract words could be used with text-to-image generators that can merge together completely unrelated terms (e.g. "love" and "web browser") and generate (meaningful) visuals from the description combining them (e.g. "love embracing the web browser"). Recall of words with low "imageability" could potentially benefit from text-to-image generated images, which should be explored in the future.

### 7.2 Limitations

One of the problems we observed during our studies was the restriction of text-to-image generators. They can not use words or a combination of words that are potentially offensive. Several user descriptions could not be converted into images, even though we did not consider them offensive in the context presented (although the words used could be employed to generate offensive content). Actually, more than offensive, they were considered bizarre or absurd.

All studies were carried out in English, which was not the first language of participants, and which could present a potential limitation. Nevertheless, all of participants were fluent in English at a high level. Our studies also show that participants generated keywords in multiple languages and describe associations in English. This highlights the importance of multiple language support in keywords based vocabulary learning systems. It also highlights one of the key limitations of the current text-to-image generators – the are limited to English language only.

Another limitation is focusing on one technique only. The keyword method is just one of many techniques that can be used to learn vocabulary of a foreign language. The favourite technique varies from learner to learner and even for the same learner over time. In addition, different learning techniques can be used and combined at the same time. It is nevertheless worth exploring the possibilities to enhance one technique only by employing novel technologies.

For the Slovenian as a target language, we had two words in each condition that are similarly pronounced in English language because we decided to use the same set of words for both target languages (Japanese, Slovenian).

This limitation only affects language assessment since all other independent variables are counterbalanced. Despite, the group learning Slovenian reported a higher workload, which again points to cultural differences in experiencing workload.

Lastly, we tested the recall twice – immediately after the study and 7 days later. While the language learning is a longitudinal process and not a two times event. Despite this, we have shown short-time benefits of viewing text-to-image generated images for the keyword method.

## 8 CONCLUSION

In the studies presented, we have used text-to-image generators to amplify the effectiveness of the keyword method for learning vocabulary. The keyword method is based on visualising a link between the meaning of the word and what its pronunciation in the foreign language reminds us of phonetically. This approach is successful due to its reliance on mental imagery to support memory retention and recall. Our findings reveal that externalising these mental visualisations into images using text-to-image generators adds visual stimuli that further enhances memory retention and recall.

The results of our studies show that users are able to describe their mental visualisations by using the keyword method. The studies revealed that users come up with keywords mainly in their native language but also used words in other languages they speak, (geographical and personal) names, or anything else that sounded like the foreign word to them. Text-to-image generators were also capable of producing images that users considered suitable. But most importantly, we have shown that by externalising mental links in visually, users were able to remember more words in immediate and delayed recall. Finally, despite the fact that images generated by DALL-E 2 were selected more often, in approximately 66% of the cases images that made it to the final selection were from at least two different models. This shows that participants preferred a more heterogeneous set of images to select from.

## ACKNOWLEDGMENTS

This research was funded by the (i) Slovenian Research Agency, grant numbers P5-0433, IO-0035, J5-50155 and J7-50096, (ii) research program CogniCom (0013103) at the University of Primorska, and (iii) by the EU EIC Pathfinder-Awareness Inside challenge "Symbiotik" project (from 1 Oct 2022 to 30 Sept 2026) under Grant no. 101071147.

Authors version.

## A APPENDIX

In Table 3 are the words we used in all three studies. The IMAEG variable is the imageability rating from the Medical Research Council (MRC) psycholinguistic database [84]. The database is available online on the following URL https://websites.psychology.uwa.edu.au/school/mrcdatabase/uwa_mrc.htm. The IMAEG variable has values in the range 100 to 700 (min 129; max 669; mean 450; s.d. 108).

We decided that the value for words selected needs to be at least 500. Some words do not exist in the database (bookshelf, bicycle, cupboard, zebra, and giraffe). We used a method for assessing 'imageability' of these words in which 'imageability' scores of words with similar meaning were taken. We used the set from Study B also as one of the sets in Study C. Each set contains two animals, two clothing or jewellery pieces (something people wear), one vehicle, one furniture, one plant, and the rest are common objects from around the house.

**Study A**

| | Word | IMAEG |
|---|---|---|
| 1 | Postcard | 578 |
| 2 | Tap | 541 |
| 3 | Wall | 575 |
| 4 | Chair | 610 |
| 5 | Pencil | 607 |

**Study B**

| | Word | IMAEG |
|---|---|---|
| 1 | Bookshelf | |
| 2 | Lamp | 575 |
| 3 | Flower | 618 |
| 4 | Radio | 613 |
| 5 | Telephone | 655 |
| 6 | Bicycle | |
| 7 | Socks | 553 |
| 8 | Shoes | 601 |
| 9 | Elephant | 616 |
| 10 | Rabbit | 611 |

**Study C**

| | Word | IMAEG | Japanese | Slovene |
|---|---|---|---|---|
| 1 | Bookshelf | (furniture has a value 588) | 棚 | Polica (knjižna) |
| 2 | Lamp | 575 | ランプ | Svetilka |
| 3 | Flower | 618 | 花 | Roža |
| 4 | Radio | 613 | ラジオ | Radio |
| 5 | Telephone | 655 | 電話 | Telefon |
| 6 | Bicycle | (vehicles usually above 550) | 自転車 | Kolo |
| 7 | Socks | 553 | 靴下 | Nogavice |
| 8 | Shoes | 601 | 靴 | Čevlji |
| 9 | Elephant | 616 | 象 | Slon |
| 10 | Rabbit | 611 | うさぎ | Zajec |
| 1 | Clock | 614 | 時計 | Ura |
| 2 | Vase | 563 | 花瓶 | Vaza |
| 3 | Mirror | 627 | 鏡 | Ogledalo |
| 4 | Trousers | 630 | ズボン | Hlače |
| 5 | Car | 638 | 車 | Avto |
| 6 | Cupboard | (furniture has a value 588) | 食器棚 | Omara |
| 7 | Ring | 601 | 指輪 | Prstan |
| 8 | Zebra | (common animals above 600) | シマウマ | Zebra |
| 9 | Tree | 622 | 木 | Drevo |
| 10 | Giraffe | (common animals above 600) | キリン | Žirafa |

Table 3. The words and their IMAEG "imageability" ratings from the MRC database.

## B  APPENDIX

The detailed results of the questionnaire measuring appeal (e.g. attractiveness, desirability), hedonic (e.g. novelty, originality), and ergonomic (e.g. simplicity, familiarity) properties [22, 23] (used in software and product evaluation) are shown in Figure 11.

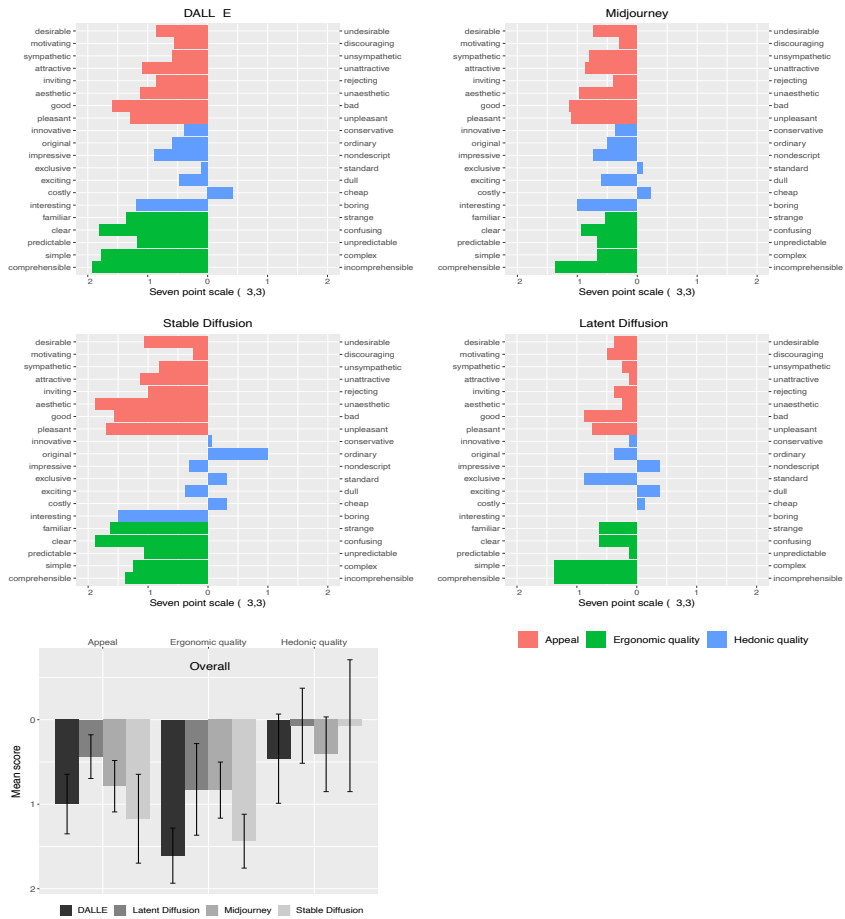

Fig. 11. Individual scores assessing appeal, hedonic qualities and ergonomic qualities of generated images for each text-to-image generator.

## C APPENDIX

In Table 4 are ten examples of keywords and associations the participants came up with in English and generated images.

| Japanese word | Rōmaji | Keyword | English Word | Association | Generated images |
|---|---|---|---|---|---|
| 本棚 | Hondana | Honda | Bookshelf | Honda driving into a bookshelf | 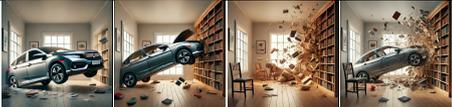 |
| 象 | Zō | Zoo | Elephant | Elephant in a cage. | 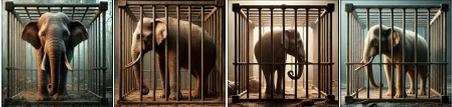 |
| 無線 | Musen | Mussels | Radio | A plate of mussels near a radio | 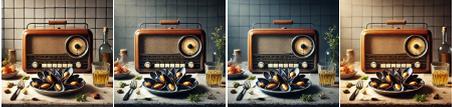 |
| 花 | Hana | Henna | Flower | A henna drawing of flowers | 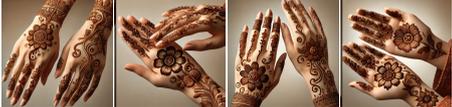 |
| 靴下 | Kutsushita | Cat's son | Socks | A bunch of socks with the design of baby cats | 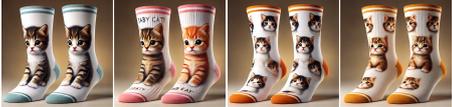 |
| 靴 | Kutsu | Stew | Shoes | A stew made out of shoes | 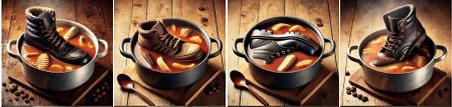 |
| ズボン | Zubon | Ribbon | Trousers | Trousers with ribbons | 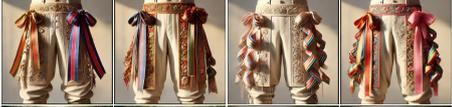 |
| 本棚 | Hondana | Honda and Ana | Bookshelf | A girl on Honda bike driving down a curved road covered with books | 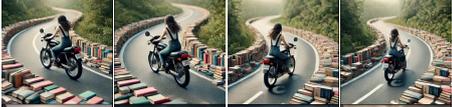 |
| 自転車 | Jitensha | Tension | Bicycle | A bike driving on a tight rope | 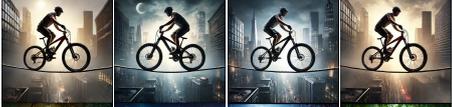 |
| うさぎ | Usagi | A saga | Rabbit | Rabbits in trouble in a fairytale | 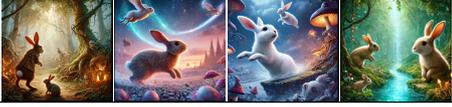 |

Table 4. Ten randomly selected words, with keywords, associations, and generated images. The rōmaji column is romanization of Japanese words.